\documentclass{ws-procs9x6}

\begin{document}

\title{Reaction Rates and Nuclear Properties Relevant for Nucleosynthesis in
Massive Stars and Far From Stability}

\author{T. RAUSCHER, C. FR\"OHLICH}

\address{Departement f\"ur Physik und Astronomie,\\
Universit\"at Basel, \\
CH-4056 Basel, Switzerland\\
E-mail: Thomas.Rauscher@unibas.ch}

\author{K.H. GUBER}

\address{Nuclear Science and Technology Division, \\
Oak Ridge National Laboratory, \\
Oak Ridge, TN 37831, USA}


\maketitle

\abstracts{
Explosive nuclear burning in astrophysical environments produces
unstable nuclei which again can be targets for subsequent reactions. In
addition, it involves a large number of stable nuclides which are not
fully explored by experiments, yet. Thus, it is necessary to be able to
predict reaction cross sections and thermonuclear rates with the aid of
theoretical models. Such predictions are also of interest for
investigations at radioactive ion beam facilities.
An extended library of theoretical cross sections and reaction rates
is presented. The problem of $\alpha$+nucleus potentials is addressed
and new parametrizations presented. The problem of properly predicting
cross sections at low level densities is illustrated by the $^{62}$Ni(n,$\gamma$) reaction.
}

\section{Introduction}
The majority of reactions in astrophysics involving
the strong interaction can be described in the statistical 
model~\cite{rtk}.
In predictions of cross sections for astrophysical applications,
slightly different points are emphasized than in pure nuclear physics
investigations. Firstly, one is confined to the very low energy region,
from thermal energies up to a few MeV. Secondly, since most of the
ingredients for the calculations are experimentally undetermined, one has
to develop reliable phenomenological or microscopic models to predict
these properties with an acceptable accuracy across the nuclear chart.
Therein one has to be satisfied with a more limited accuracy as compared
to usual nuclear physics standards. Considering the substantially
larger uncertainties in many astrophysical scenarios, this seems to be
adequate.

\section{Statistical model calculations}
A recently published large-scale reaction rate library includes 
neutron-, proton-, and $\alpha$-induced reactions on all target
nuclei from Ne up to Bi from proton-dripline to neutron-dripline~\cite{rt00,rt01}.
Due to the fact that many very short-lived nuclides can be produced
in astrophysical sites, it is necessary to provide cross sections and
rates for about 4600 targets and 32000 reactions. These numbers show
that theory will always play a major role in providing cross section,
despite the potential of future Rare Isotope Accelerators. The calculations
were performed with the Hauser-Feshbach code NON-SMOKER~\cite{nonsm} which
is especially tuned to such large-scale predictions. Details of the
nuclear properties used are given elsewhere~\cite{rt00}. This rate set has
already be adopted as a standard for nucleosynthesis in stellar evolution
and in type II supernovae~\cite{snii}.

Fits to the astrophysical reaction rates -- ready for direct astrophysical
application -- as well as tables of cross sections, reaction rates, and
nuclear inputs for all possible reactions with light projectiles
can also be downloaded from {\it http://nucastro.org/reaclib.html} .

\section{Optical $\alpha$-nucleus potentials}
There have only been few attempts to derive global optical potentials
for $\alpha$-projectiles~\cite{raunic} and most of them are only valid at 
$\alpha$-energies
larger than 30 MeV. Due to the high Coulomb barrier and nuclear structure
effects defining the imaginary part of the potential it is difficult to
obtain a global potential at astrophysical energies. Elastic $\alpha$-scattering
data can constrain the real part of the potential~\cite{mcf,mohr97} and
detailed analysis can also improve on the imaginary part~\cite{mohr00,molyb}, 
describing the absorption into other channels than the elastic scattering, 
i.e. the Hauser-Feshbach channel. Due to the scarcity of data for intermediate
and heavy nuclei, attempts to improve on the potential are mostly
concentrating on single reactions~\cite{endre,molyb}. More global
approaches suffer from the lack of data to confine their parameters~\cite{raunic,gornic}.

We have tried to find a potential for the $A\equiv 140$ mass region
by simultaneously fitting
data for $^{143}$Nd(n,$\alpha$)$^{140}$Ce~\cite{koehpriv}, 
$^{147}$Sm(n,$\alpha$)$^{144}$Nd~\cite{koeh}, and
$^{144}$Sm($\alpha$,$\gamma$)$^{148}$Gd~\cite{endre}.
The optical potential is parametrized as
\begin{equation}
V(r,E)=-\frac{V_0}{1+\exp\left(\frac{r-r_rA^{1/3}}{a_r}\right)}
-i\frac{W(E)}{1+\exp\left(\frac{r-r_VA^{1/3}}{a_V}\right)} \quad.
\end{equation}
Different
parameters for the potential geometry and the energy dependence of the
depth of the imaginary part were explored~\cite{froh}. We did not find
significant differences between using a Brown-Rho shape~\cite{mohr97} 
$W(E)=W_0 ((E-E_0)^2)/((E-E_0)^2+\Delta^2)$ or a
Fermi-type shape~\cite{endre} $W(E)=W_0 / (1+\exp((E^*-E)/a^*))$
of the energy dependence. For the latter
we found the parameters $E^*=18.74$ MeV, $a^*=2.1$ MeV, with all other 
parameters
as in the previous paper~\cite{endre}. The Brown-Rho best fit was obtained
with $E_0=6.35$ MeV and $\Delta=28.4$ MeV, with the same fixed
parameters $V_0=162$ MeV, $r_r=1.27$ fm, $a_r=0.48$ fm, $W_0=19$ MeV, $r_V=1.57$
fm, $a_V=0.6$ fm. The results from the
simultaneous fit of three reactions are shown in Figs.\ 1 and 2.
\begin{figure}[t]
\begin{center}
\epsfxsize=11cm
\epsfbox{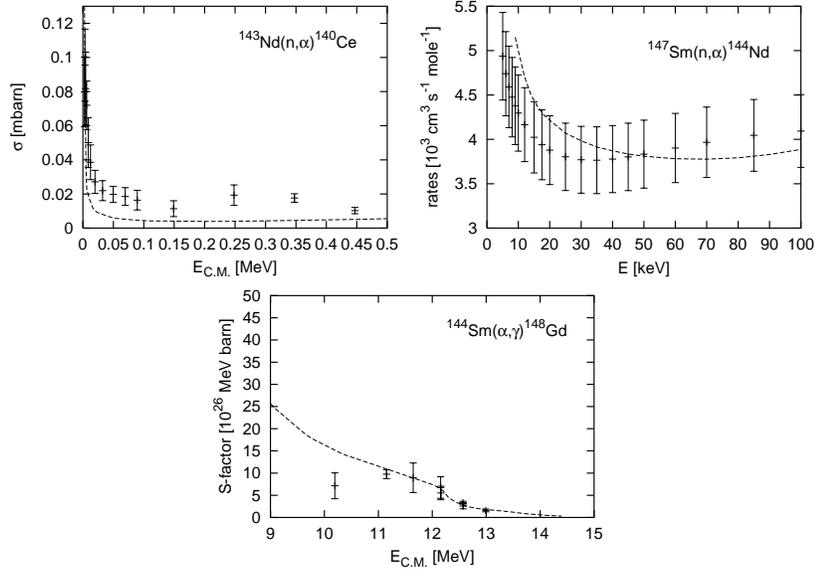}
\end{center}
\caption{Cross sections, reaction rates, and S-factors from a
simultaneous $\chi^2$ fit of the Fermi-type energy-dependent $\alpha$+nucleus
optical potentials of three reactions (see text). The dashed lines are
the statistical model calculation. The errors on the
$^{147}$Sm(n,$\alpha$) rates were assumed to be 10\%.}
\end{figure}
\begin{figure}[t]
\begin{center}
\epsfxsize=11cm
\epsfbox{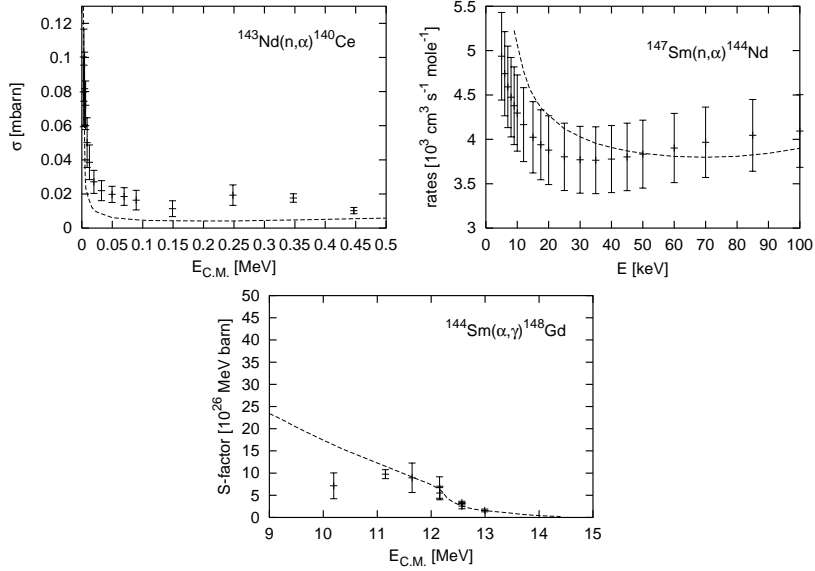}
\end{center}
\caption{Cross sections, reaction rates, and S-factors from a
simultaneous $\chi^2$ fit of the Brown-Rho energy-dependent $\alpha$+nucleus
optical potentials of three reactions (see text). The dashed lines are
the statistical model calculation. The errors on the
$^{147}$Sm(n,$\alpha$) rates were assumed to be 10\%.}
\end{figure}

Despite the fact that the considered targets are in the same mass region,
the derived parameters also describe acceptably well the reaction
$^{96}$Ru($\alpha$,$\gamma$)$^{100}$Pd~\cite{ru}. However, it is
remarkable that even better overall agreement with all four reactions
can be obtained when using a mass- and energy-independent
potential of Saxon-Woods form for the real and 
imaginary parts (see Fig.\ 3). The real parameters are given by $V_0=162.3$ MeV,
$r_r=1.27$ fm, $a_r=0.48$ fm, the imaginary ones by $W_0(E)=W_V=25$ MeV, 
$r_V=1.4$ fm,
$a_V=0.52$ fm. Thus, the real part is identical to the potential by
Somorjai {\it et al.}~\cite{endre} but without energy dependence, 
whereas the imaginary part is similar
to the one used in McFadden \& Satchler~\cite{endre}.
Since the McFadden \& Satchler parameters were
derived from extensive elastic scattering data it seems reasonable that
they are applicable to a wider range of targets. The Somorjai
{\it et al.} parameters were derived for one reaction only
but seem to work also for the nuclides investigated here. Certainly, 
at very low
$\alpha$-energies an additional energy-dependence has to be introduced.
Here, we do not show our results from fitting each reaction separately.
Obviously, potentials fitted to single reactions can describe those --
but only those -- even better.
\begin{figure}[t]
\begin{center}
\epsfxsize=11cm
\epsfbox{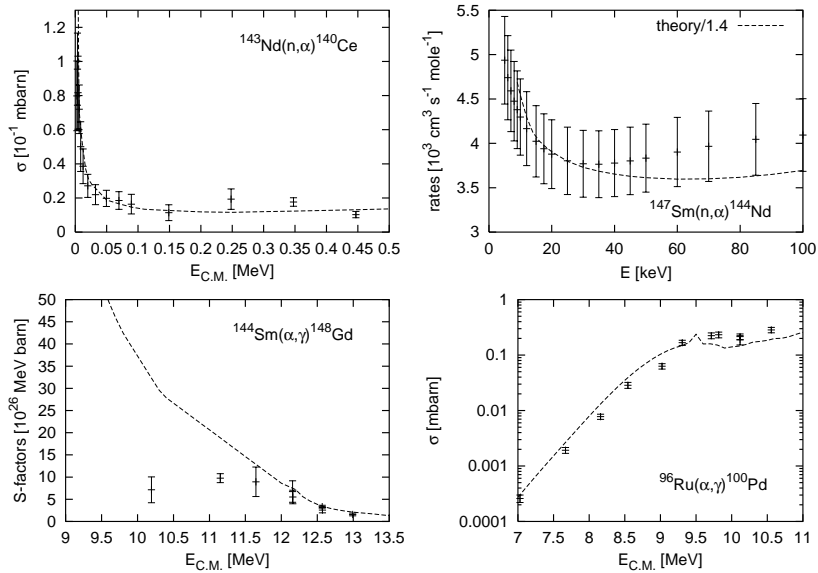}
\end{center}
\caption{Results for four different reactions using the energy-independent
potential (see text). The dashed lines are the statistical model
calculation.
Note that the $^{147}$Sm(n,$\alpha$) result is
renormalized by a factor 1/1.4.}
\end{figure}

\section{The $^{62}$Ni(n,$\gamma$) case}
For neutron-induced reactions at low energies,
close to magic numbers, and far off stability where low separation
energies are encountered, another problem emerges. In such targets, the
level density is too low to allow the application of the statistical
model~\cite{rtk}. Also for other nuclides it is not straightforward to
bridge the region of thermal energies to the region of overlapping
resonances where the Hauser-Feshbach formalism can be used. Single resonances
and direct reactions become important. This is also an issue for neutron-rich
nuclei in the $r$-process path with low neutron-separation energies.

As an example for the difficulties in extrapolating thermal data to
$s$-process energies of up to a few hundred keV, the reaction
$^{62}$Ni(n,$\gamma$)$^{63}$Ni is discussed here. Two compilations give
disagreeing 30 keV cross sections~\cite{bao,bao00}, based on the same 
thermal data. Both extrapolations assume s-wave behavior of a direct
capture component. The more recent one includes a sub-threshold resonance
contributing to the thermal cross section.

We have calculated the direct capture component using DWBA and found a
considerable p-wave contribution which enhances the cross section at
30 keV~\cite{guber}. Thus, even when including the subthreshold
resonance, the 30 keV value is coincidentally similar to the value in the
older compilation (Fig.\ 4). However, also the general energy dependence of the
cross section is altered. Resonances were also included but
they only contribute less than 15\%. The enhanced cross section
has an important impact on $s$-processing in massive stars. A previously
seen overproduction of $^{62}$Ni in stellar models can be cured when using our
enhanced rate because of increased destruction of this nucleus with
the larger neutron capture rate~\cite{snii,guber}.
\begin{figure}[t]
\begin{center}
\epsfxsize=11cm
\epsfbox{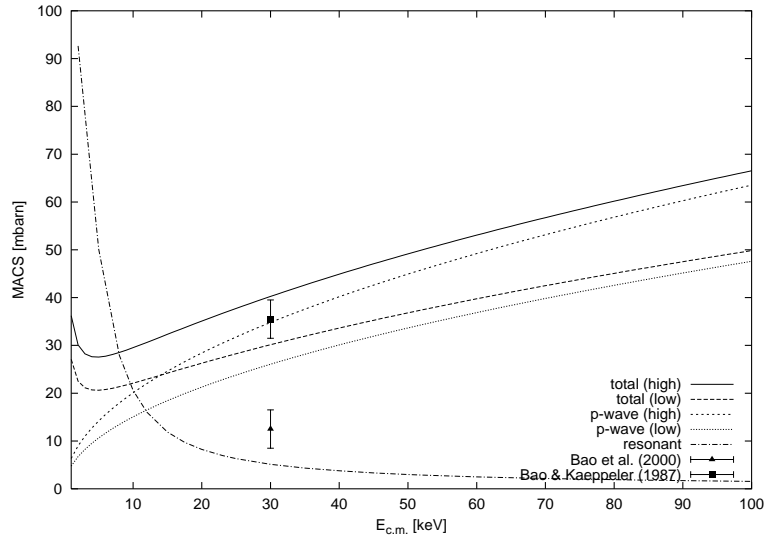}
\end{center}
\caption{Direct neutron capture Maxwellian averaged cross section of
$^{62}$Ni. The final value is given by adding the resonant contribution
to the ``total'' direct term. Upper and lower limits on the direct
components are from experimental errors on the input, {\it i.e.} in the
thermal scattering length and the spectroscopic factors.}
\end{figure}

\section{Conclusions}
Despite considerable successes in the prediction of cross sections and
reaction rates close to and far off stability, the description of
certain nuclear inputs, such as optical $\alpha$-potentials, still needs
to be improved. It is also still unclear whether nuclear properties
far off stability can be predicted with sufficiently high accuracy.
Although future advances in microscopic theories may alleviate that
problem, experimental data is clearly needed. Rare Isotope Accelerators
will make it possible to study highly unstable nuclides but also
``classical'' nuclear physics experiments with stable or long-lived
nuclei are indispensable. They can provide the systematics for global
descriptions and shed light on the interaction of different reaction
mechanisms.

\section*{Acknowledgements}
This work was supported in part by the Swiss NSF grants 2124-055833.98,
2000-061822.00, 2024-067428.01) and the U.S. DOE.
T.R. acknowledges a PROFIL professorship from the Swiss NSF.

\end{document}